\documentclass[11pt]{article}
\usepackage{moriond,epsfig}

\def\lsim{\stackrel{<}{\sim}}

\def\wm{w_Q^{m}}
\def\w0{w_Q^{0}}
\def\at{a_c^{m}}
\def\s8{\sigma_8}

\def\be{\begin{equation}}
\def\ee{\end{equation}}
\def\bea{\begin{eqnarray}}
\def\eea{\end{eqnarray}}

\begin{document}
\vspace*{4cm}
\title{Dark Energy in Chains}

\author{David Parkinson${}^1$, Bruce A.~Bassett${}^{1,2}$, Edmund J.~Copeland${}^3$, Pier-Stefano Corasaniti${}^{4}$ \& Martin Kunz${}^5$ }

\address{${}^1$ Institute of Cosmology and Gravitation, University of
  Portsmouth, Portsmouth, PO1 2EG, UK \\
  ${}^2$ Department of Physics, Kyoto University, Kyoto, Japan\\
  ${}^3$ Department of Physics and Astronomy, University of Sussex,
  Brighton, BN1 9QJ, UK \\
  ${}^4$ ISCAP, Columbia University, Mailcode 5247,
  New York NY 10027, United States \\
  ${}^5$ Astronomy Centre, University of Sussex,
  Brighton, BN1 9QJ, UK}

\maketitle\abstracts{Dark energy affects the CMB through its perturbations 
and affects both CMB and Sn Ia through its background evolution. 
Using recent CMB and Sn Ia data sets, together with the most general 
parametrization of the dark energy equation of state available,
we find that today $w<-0.8$ ($2\sigma$). We also find that 
the value of the normalization of the power spectrum on cluster scales, 
$\sigma_8$, can be used to discriminate between dynamical models of 
dark energy (Quintessence models) and a cosmological constant model 
($\Lambda$CDM).}

\section{Introduction}

The WMAP satellite measurements of the Cosmic Microwave Background 
anisotropies \cite{BENNETT} have provided accurate determinations of 
many of the fundamental cosmological parameters.  When combined with 
other data sets such as the luminosity distance to type-Ia supernovae 
or large scale structure (LSS) data \cite{PEL,TONRY,KNOP,ef}, they 
reinforce the need for an exotic form of dark energy, which is 
characterized by a negative pressure and is responsible for the observed
accelerated expansion of the universe.  There are two main scenarios 
used to explain the nature of the dark energy, a time independent
cosmological constant $\Lambda$, an evolving scalar field 
(Quintessence) \cite{WETTERICH,ZATLEV,AMS1}.  Previous tests of 
quintessence with pre-WMAP CMB data \cite{CORAS1,MORT,BRUCE}, have led to 
constraints on the 
value of the dark energy equation of state parameter, $w_Q \lsim -0.7$ with 
the cosmological constant value, $w_\Lambda=-1$ being the best fit.  
Nevertheless a dynamical form of dark energy is not excluded.  Specifically 
the detection of time variation of $w$ would be of immense importance as 
it would rule out a simple cosmological constant scenario.

We perform a model independent analysis of the time evolution of the
dark energy equation of state.  We conduct the likelihood analysis using 
the WMAP data \cite{BENNETT} and the Sn Ia luminosity distance 
data \cite{PEL,TONRY}.

\section{Method and Data}

We parametrize the equation of state $w$ using five
dark energy parameters ($\overline{W}_{Q}$). They are: the value of $w$ today,
$\w0$, its value at high redshift, $\wm$, the value of the scale
factor where $w$ changes between these two values, $\at$ and the
width of the transition, $\Delta$. We are using the form
advocated in Corasaniti \& Copeland \cite{CORAS2}, which has been shown 
to allow adequate
treatment of generic quintessence and to avoid the biasing problems
inherent in assuming that $w$ is constant.

We also include the cosmological parameters
$\overline{W}_{C}=(\Omega_Q,\Omega_b h^2,h,n_S,\tau,A_s)$,
which are the dark energy density, the baryon density,
the Hubble parameter, the scalar spectral index,
the optical depth and the overall amplitude of the fluctuations
respectively.  We are assuming a flat universe. 
We therefore end up with ten parameters which can be
varied independently.

There is a degeneracy in $n_S$, $\tau$ and $\Omega_b h^2$,
which allows the models to reach unphysically high values of the
baryon density and the reionisation optical depth. 
Following the WMAP analysis we place
a prior on the reionisation optical depth, $\tau \leq 0.3$.
We also limit ourselves to models with $w(z) \geq -1$.

In order to compute the CMB power spectra, we use a modified 
version of the CMBfast Boltzmann solver \cite{ZALDA}.  Rather 
than using grid-based analysis  (which would necessitate very coarse
sampling), we opted for a Markov-Chain Monte Carlo (MCMC) approach. 
We ran 16 to 32 independent chains on the UK national cosmology supercomputer
(COSMOS). This approach has both the advantage that there was no need
to parallelize the Boltzmann solver, and lets us assess the convergence and
exploration by comparing the different chains.

\section{Results}

\begin{figure}[h]
\begin{center}
\includegraphics[width=70mm]{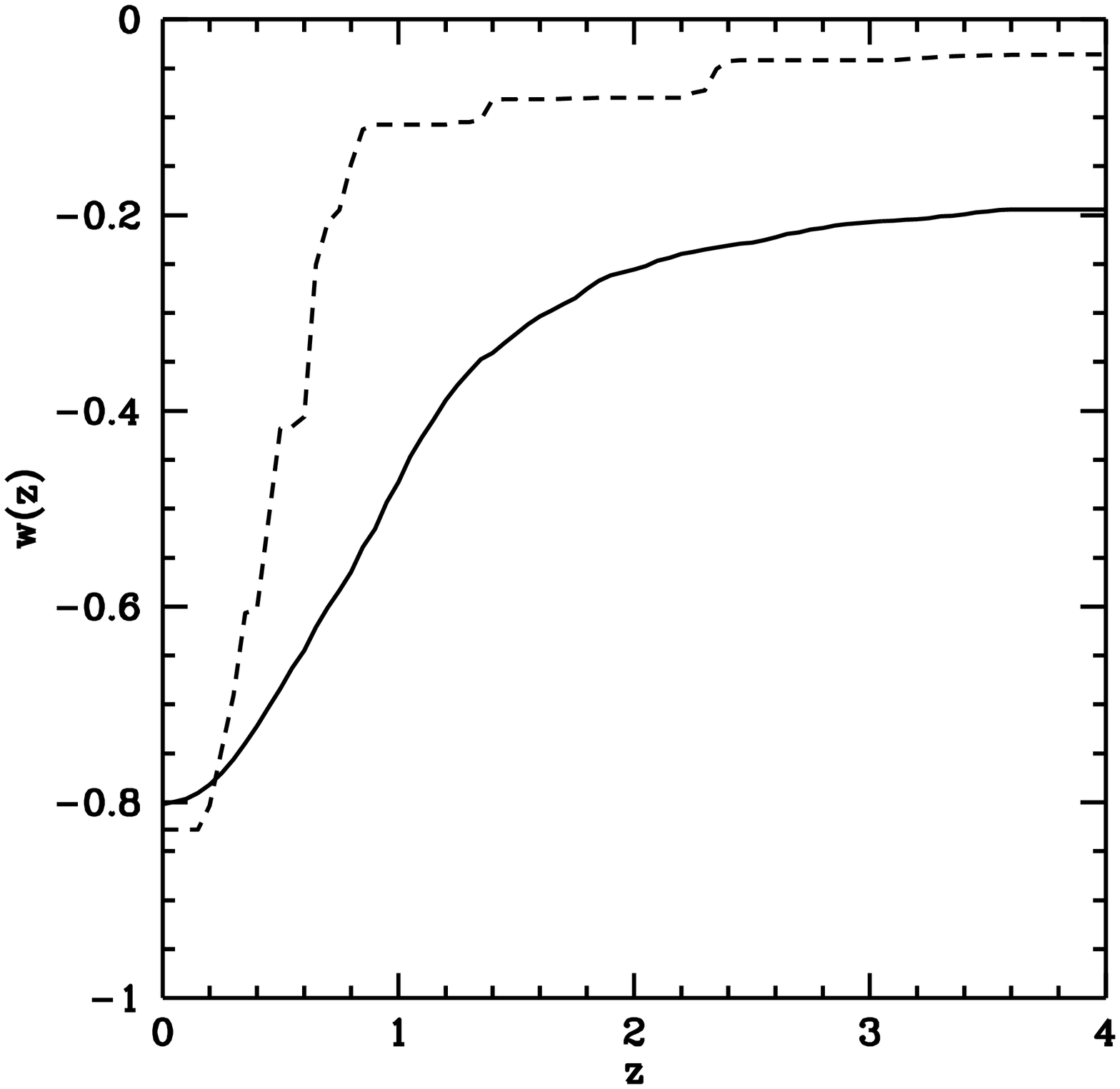} 
\includegraphics[width=70mm]{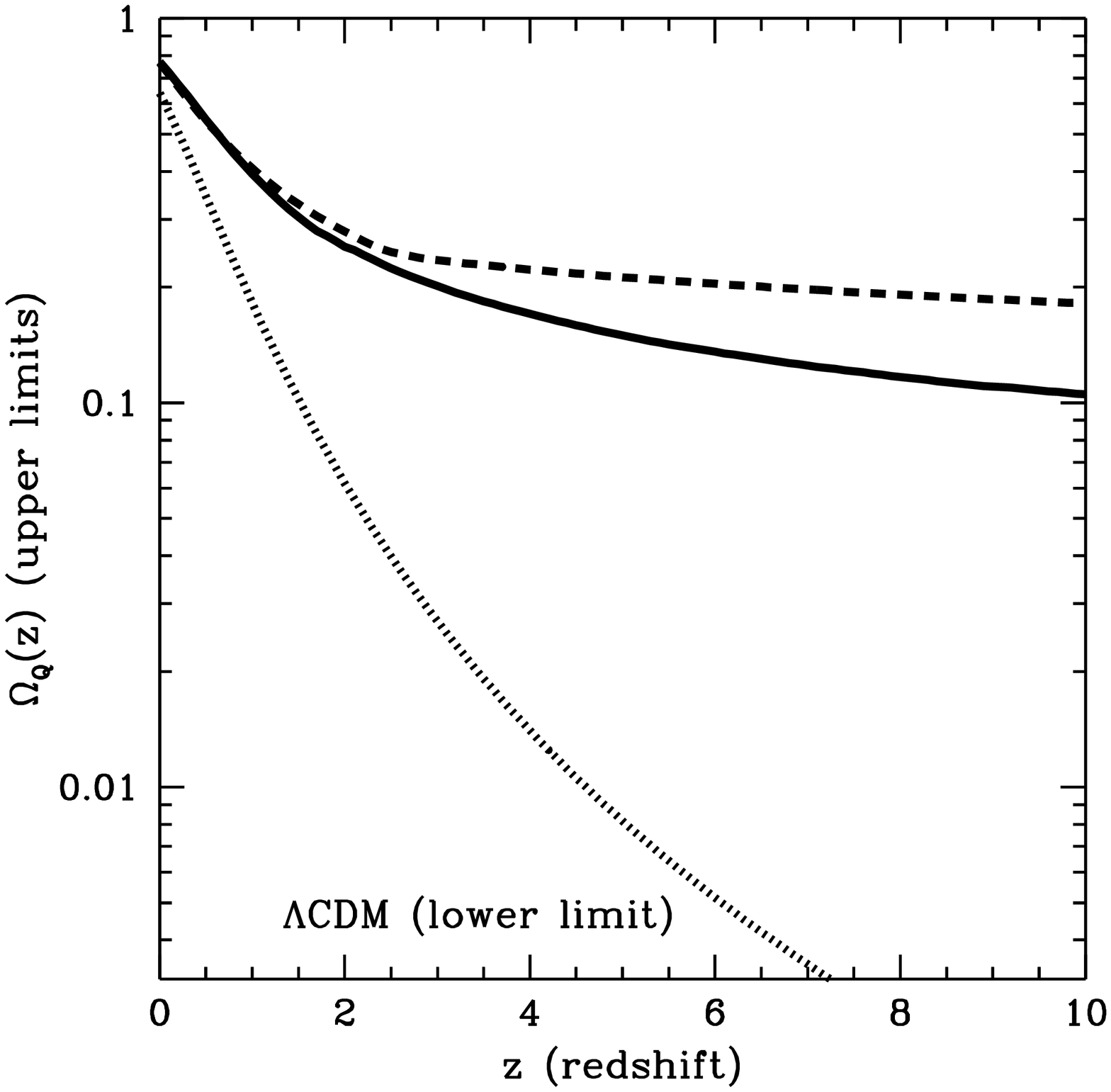} \\
\caption[wz]{\label{wz} Upper $2\sigma$ limits on $w(z)$ (left) and
$\Omega_Q(z)$ (right) derived by taking 
the 95\% models with lowest $w(z)$ from our main chain (solid), 
searching for the highest $w(z)$ for models with $\Delta-\chi^2 < 4$ from 
the best-fit model in our main chain (dashed).  $\Lambda$CDM is acceptable
at $2\sigma$ and so there is no lower limit on $w(z)$.}
\end{center}
\end{figure}

Our global best fit QCDM model has the dark energy parameters $\w0=-0.99$,
$\wm=-0.11$, $\at=0.50$ and $\Delta=0.079$, which corresponds to a fast 
transition at redshift of $1$  The total $\chi^2$ of the model
is $1604$, compared to the $\Lambda$CDM model $\chi^2=1606$.  The number of
degrees of freedom is $1514$, so all our fits are bad, but this is mainly due
to issues with the WMAP data (see the discussion in Spergel {\it et al.} \cite{SPERGEL}).

The WMAP CMB data constrains the cosmological parameters
$\overline{W}_{C}$ in a range of values consistent with
the results of previous analysis \cite{SPERGEL,LUCA,DORAN}.
The addition of the dark energy parameters $\overline{W}_{Q}$ does not 
introduce any new degeneracies with the other parameters.  However,
there are new internal degeneracies between the dark energy parameters.
In particular the only parameter we can constrain well is the
equation of state today $\wm < 0.8$ at $2\sigma$.  A more complete
discussion of this will be available in a forthcoming paper \cite{DELONG}.

\begin{figure}
\begin{center}
\psfig{figure=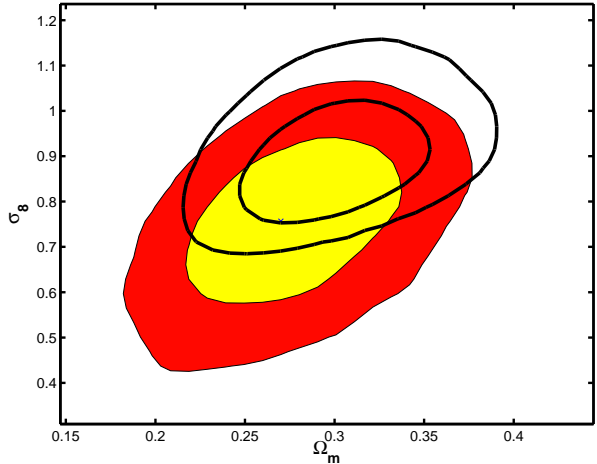,width=80mm}
\caption[oms8]{\label{oms8} Marginalized 68\% and
95\% confidence contours for quintessence (filled contours) and
$\Lambda$CDM models (solid lines). $\Lambda$CDM
has a systematically higher value of $\sigma_8$, and
a slightly higher value of $\Omega_m$.}
\end{center}
\end{figure}

\begin{figure}
\begin{center}
\psfig{figure=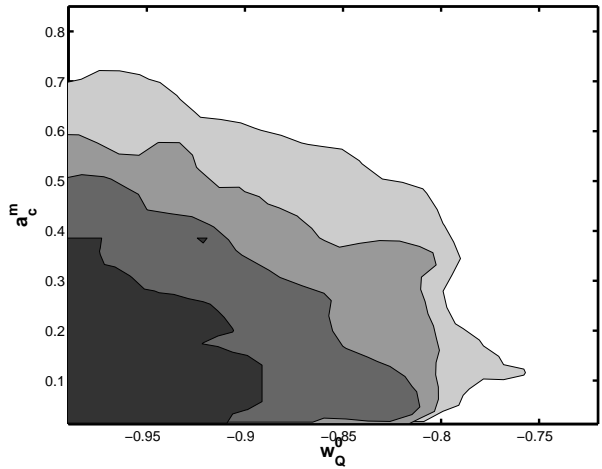,width=80mm}
\caption[atw0]{\label{atw0} The 95\% confidence regions  
for models with a rapid transition for different limits on $\sigma_8$,
going from (lightest grey to darkest) all data, $\s8>0.75$, $\s8>0.9$ and 
$\s8>1.05$.}
\end{center}
\end{figure}

However, we have found that information about the power spectrum on cluster
scales ($\s8$), would allow us to break this degeneracy.  In Corasaniti 
{\it et al.} \cite{CBUC} it was shown the different quintessence models 
leave a different imprint on the CMB power spectrum.  Models with a more 
rapid transition at smaller 
redshifts will produce a larger ISW effect than $\Lambda$CDM.  This means 
they require a smaller value of $A_s$ to fit the CMB data, and so will have
a smaller $\sigma_8$.  Figure \ref{oms8} shows us that an independent 
measurement of $\sigma_8$ would allow us to distinguish between $\Lambda$CDM
and a time dependent dark energy component.

This is shown in more detail in fig.~\ref{atw0}.  Here we plot the 95\%
confidence regions for rapid transition models ($\wm > -0.3$ and 
$\at/\Delta > 1.2$ which includes our best fit model) with different 
limits on the value of $\s8$.  $\Lambda$CDM 
will correspond to $\w0=-1$ and $\at \rightarrow 0$, and so will sit in the 
bottom left-hand corner of this plot, favouring high-$\s8$ models.  As we move
away from this corner, the limit on $\s8$ falls.  If we restrict ourselves
to models with high-$\sigma_8$ we favour $\Lambda$CDM-similar models, 
while in the opposite case we can exclude them.  For more
discussion on this area see our previous paper \cite{clustering}.

\section{Conclusions}

We have analyzed the dark energy with a model-independent approach using
CMB and Sn Ia data.  We have found that of our 4 dark energy parameters, 
only the equation of state today is well constrained, with $\w0 < -0.80$.  
We also see no strong change in $w$ for $z<1$. There are no new degeneracies
between our extra dark energy parameters and the other cosmological
parameters.  The degeneracies in the dark energy parameters may be broken
using clustering data, which could ultimately be used to distinguish QCDM 
and $\Lambda$CDM.  Nevertheless, there is no significant improvement over 
$\Lambda$CDM model.

\section*{Acknowledgments}

MK and DP are supported by PPARC.  BB is supported in Kyoto by the JSPS.
We acknowledge extensive use of the UK National Cosmology Supercomputer funded
by PPARC, HEFCE and Silicon Graphics / Cray Research.

\end{document}